\documentclass[12pt]{article}

\usepackage[margin=2.5cm]{geometry}

\usepackage{graphicx}
\usepackage{tikz}
\usetikzlibrary{positioning,arrows.meta}
\usepackage{authblk}
\definecolor{centcol}{HTML}{E2571E}   
\definecolor{gamecol}{HTML}{4D59D6}   
\definecolor{conccol}{HTML}{9B59B6}   
\definecolor{flowcol}{HTML}{4C6EF5}   
\definecolor{optcol}{HTML}{E67E22}    
\definecolor{hybcol}{HTML}{2ECC71}    
\usepackage{float}
\usepackage{algorithm}
\usepackage{algorithmic}
\usepackage{amsmath,amssymb}
\usepackage{xurl}            
\usepackage{subcaption}
\usepackage{hyperref}
\usepackage{natbib}
\usepackage{lineno}
\usepackage{dcolumn}
\usepackage{multirow}
\usepackage{units}
\usepackage{booktabs}
\usepackage[table]{xcolor}
\usepackage{rotating}
\usepackage{epstopdf}        
\usepackage{authblk}         
\usepackage{caption}
\newcolumntype{.}{D{.}{.}{-1}}
\newcolumntype{d}[1]{D{.}{.}{#1}}
\usepackage{hyperref}
\hypersetup{
  breaklinks=true,
  colorlinks=true,
  citecolor=blue,
  urlcolor=blue,
  linkcolor=black
}

\title{\bfseries Power and Control in Complex Networks:\\ A Taxonomy and Critical Review}

\author[1]{A.~Abeltino}
\author[1]{T.~Bacaloni}
\author[1]{A.~Bernardini}
\author[1]{F.~Giancaterini\footnote{Corresponding author, email: fgiancaterini@fub.it}}
\author[1]{A.~Pannone}

\affil[1]{\small Fondazione Ugo Bordoni, Italy.}

\date{\today}

\begin{document}
\maketitle
\begin{abstract}
\noindent This paper reviews the main network analysis methods used to measure structural power, which refers to the ability to shape outcomes through network position and influence, and the ability to affect others through network connections. These approaches have been applied in fields such as corporate control, global value chains, and technology supply networks. Despite significant advances, a unified framework that systematically connects these methodologies to their conceptual foundations has yet to emerge. To fill this gap, the paper introduces a taxonomy that categorizes existing methods into six families: centrality-based approaches, game-theoretic models, concentration measures, flow-based methods, optimization frameworks, and hybrid approaches that combine elements from different approaches. This classification clarifies their assumptions, analytical focus, and relative strengths, offering a coherent view of how power is structured and transmitted in complex economic and political systems. The paper concludes by outlining future research directions to refine hybrid models linking decision-making and network flows.
\end{abstract}

\renewcommand{\thefootnote}{\arabic{footnote}}

\bigskip


\section{Introduction}
Rising geopolitical tensions are reshaping the global distribution of resources and influence, as supply chains reorganize around trusted partners and cross-block exchanges become increasingly restricted; see \cite{yucesan2025does} and \cite{merlevede2024home}. The outcome is not only a shift in the scale of flows, but also the emergence of a new global structure in which interdependencies are reorganized and vulnerabilities redistributed (\cite{lopez2025modeling}). In this context, aggregate indicators (such as total trade volumes, capital flows, or market shares) tend to obscure these structural transformations and do not capture how power and influence are being reconfigured across supply chains, financial systems, political arenas, and capital markets. For example, a country that once diversified its exchanges between partners B and C may now rely exclusively on B. The aggregate figures might remain similar, but the distribution of vulnerabilities and the locus of power are fundamentally transformed. This contrast underscores both the limits of aggregate statistics and the analytical value of network approaches. Network analysis, in fact, provides a structural lens that reveals what aggregate measures often conceal: the role of critical hubs, brokerage positions that connect otherwise disconnected blocs, and the emergent architectures that shape systemic resilience and fragility. In this sense, measures of centrality and intermediation illustrate how actors derive influence not only from volume but also from their structural position within the network.

Over the years, the relevance of these models has been demonstrated in multiple domains. In corporate ownership and control, they reveal that even without majority stakes, shareholders can exercise disproportionate influence through ownership chains, extending control beyond their direct equity shares (\cite{vitali2011network}). Applied to capital centralization, network methods show how interconnected holdings translate into systemic dominance, exposing patterns of control invisible to traditional concentration ratios; see \cite{brancaccio2018centralization}. In global supply chains, they demonstrate that resilience and fragility depend less on the size of the firm than on the structural position of key hubs that transmit or absorb shocks (\cite{inoue2019firm}). In the political sphere, they highlight how the brokerage and access to information enable actors (such as states, international organizations, or NGOs) to exert influence beyond formal authority and serve as conduits for the diffusion of global standards and rules (\cite{hafner2009network} and \cite{hafner2010centrality}).

Despite this wide range of applications, the literature remains fragmented in how it defines and measures structural relationships within networks. This paper addresses this gap by proposing a taxonomy that systematically organizes network-based methodologies according to the conceptual mechanisms through which they define and measure structural power and influence, understood respectively as the capacity to shape outcomes through network position and to affect others’ decisions through network connections. The taxonomy provides a comprehensive overview of the field and enables a comparative evaluation of the assumptions, strengths, and limitations of each approach. In particular, it assesses how different methodologies perform along two critical dimensions: ultimate controllers (UO), which refers to who ultimately controls decision-making at the end of ownership chains, and intermediary power (IP), which reflects the influence of actors that transmit or mediate control between ultimate owners and target firms. We identify six main families of methods: centrality-based measures, game-theoretic models, concentration measures, flow-based methodologies, optimization approaches, and hybrid methods.

The rest of the paper is organized as follows. Section \ref{sec:res_quest} outlines the research question and the methodological approach that guide our taxonomy. Sections \ref{sec:Centrality}–\ref{sec:Optim} review the main methodological families, namely, centrality-based, game-theoretic, concentration, flow-based, and optimization measures. Section \ref{sec:Hybrid} discusses hybrid approaches that combine features of multiple frameworks to overcome the limitations of individual methods. Section \ref{sec:Conclusions} concludes with comparative information and directions for future research.

\section{Research Question and Methodological Approach}
\label{sec:res_quest}
The growing diversity of approaches has not only expanded the analytical scope of the field but also hindered the development of a unified conceptual framework. To bring analytical coherence to the study of network power, this work addresses the following research question: \textit{How can network analysis methodologies for measuring structural power be systematically classified based on their underlying conceptual mechanisms?} 

To address this question, we conducted a critical and concept-driven review of the literature. The analysis followed an iterative mapping process, beginning with the main methodological contributions that introduced different ways of conceptualizing structural power: 
\cite{freeman1978centrality} and \cite{bonacich1987power} for 
positional influence through centrality; \cite{shapley1954method} and \cite{johnston1978measurement} for pivotal power in game-theoretic 
models; \cite{hirschman1980national} and 
\cite{brancaccio2018centralization} for concentration-based measures; and \cite{vitali2011network} for flow-based formulations. Building on 
these seminal frameworks, we expanded the review to include subsequent studies that refine, combine, or operationalize these approaches in domains such as corporate control, global supply chains, and political networks. Each methodological contribution was examined through a comparative analysis of its theoretical assumptions, mathematical formulation, and analytical logic to identify the fundamental questions each approach addresses about the nature of structural power. We focus on studies that conceptualize power, control, or influence as 
relational or systemic features of networks and provide operational measures or models. This selection results in a representative body of work published mainly over the past two decades in economics, network science, and political economy.\footnote{We deliberately exclude models such as random walks and community detection (\cite{masuda2017random}; \cite{bendahman2024unveiling}), which focus on diffusion dynamics or mesoscale clustering rather than on the mechanisms that generate and transmit control and power within networks.}

\begin{figure}
    \centering
    \includegraphics[width=1\textwidth, height=0.4\textheight]{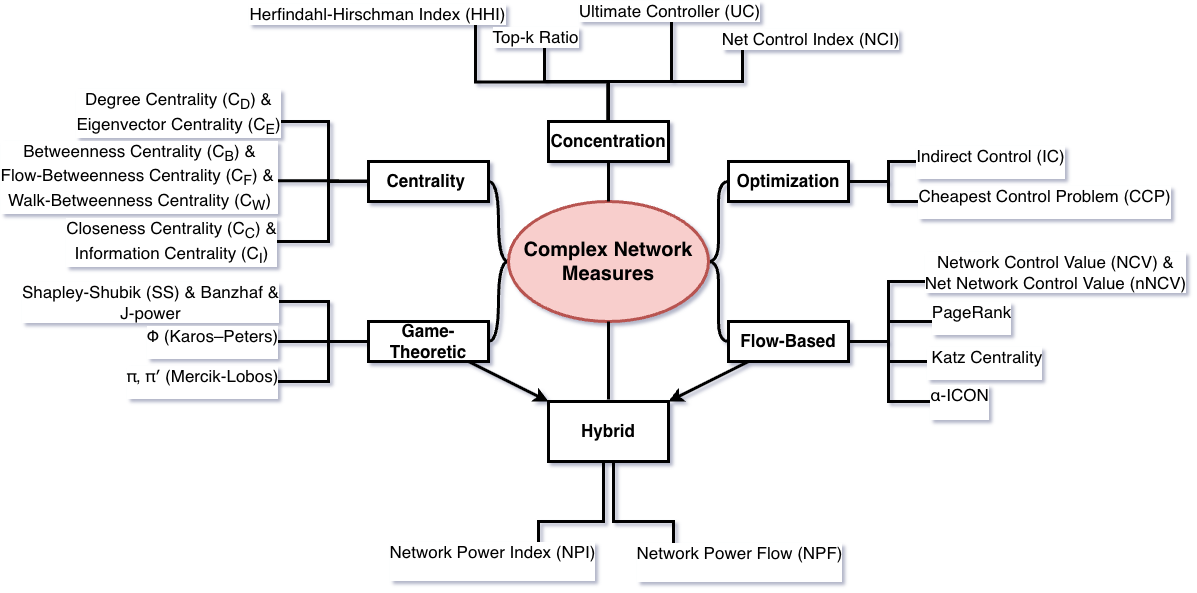}
    \caption{\textit{The figure summarizes the six main methodological families used to measure structural power and control in complex networks.}}
    \label{fig:placeholder}
\end{figure}

This analysis yields a classification based on six primary conceptual mechanisms, summarized in Figure \ref{fig:placeholder}. Centrality-based measures conceptualize influence as a function of topological position, evaluating the importance of an actor in terms of its access, brokerage role, and ability to reach others efficiently. Game-theoretic approaches interpret influence as a pivotal decision-making capacity, measuring the probability that an actor is decisive in forming a winning coalition. Concentration measures adopt a systemic perspective, interpreting power as the degree to which resources or control are concentrated and unequally distributed within the network. Flow-based measures conceptualize power as a resource that propagates through the links of a network, capturing how influence is transmitted and accumulated along relational chains. Optimization-based measures adopt a strategic perspective, defining power as the result of a cost–benefit optimization process aimed at achieving specific control objectives. Finally, hybrid methods combine elements from game-theoretic and flow-based measures to integrate decisional and propagation dynamics, linking the probability of control with the transmission of influence across the network. 

The following sections examine each of these conceptual families in detail. It is worth noting that recent studies have extended classical approaches by incorporating multi-attribute and machine-learning techniques for identifying influential nodes. Among these, \cite{sheikhahmadi2022multi} and \cite{zhao2020machine} illustrate how data-driven and hybrid models can complement traditional structural analyzes. These advances reflect an emerging convergence between network theory and artificial intelligence. They are not included in the present taxonomy, as these approaches are still in an early and evolving stage of development, representing promising directions that extend beyond the consolidated structural methodologies reviewed here.

\section{Centrality Measures}
\label{sec:Centrality}
Centrality measures are mathematical heuristics designed to identify the most important elements within a network by relying exclusively on its topology (for example, the number of direct links, the distances between nodes, or the presence of clusters and bridges). In other words, they assess the importance not of the intrinsic attributes of actors, such as wealth, size, or authority, but only of their relational configuration; see \cite{freeman1978centrality}.

As explained by \cite{hafner2010centrality}, there is no single way to measure centrality. Each measure focuses on a different type of advantage that comes from the position of a node on the network: access, brokerage, and efficiency. Access refers to having many or strong direct links, which reduces reliance on others and increases visibility. Brokerage refers to acting as a bridge between groups or actors that are not otherwise connected; this position enables a node to control, filter, or facilitate the exchange of resources and information across different parts of the network. Efficiency refers to being close to all other actors in the network; nodes in this position can reach others through fewer steps, allowing them to obtain information quickly and spread it more effectively without relying heavily on intermediaries. Together, these advantages show how the topology of a network creates structural power.

\subsection{Access Centrality Measures}
The \textit{Degree Centrality} ($C_D$) measure proposed by \cite{freeman1978centrality} is the simplest and most intuitive index of the importance of a node in a network, since it directly quantifies the power of access. Formally, it is defined as:
\begin{equation*}
\label{eq:centrality}
    C_D(n_i) = \sum_{j=1}^{N-1} a_{ij},
\end{equation*}
where $n_i$ denotes the node $i$, $N$ is the total number of nodes in the network, and $a_{ij}$ is the element of the adjacency matrix that indicates the relationship between the nodes $i$ and $j$. In weighted networks, where the edges carry a numerical value representing the strength, intensity, or capacity of a connection, $a_{ij}$ denotes this weight, for example, the percentage of ownership or the volume of trade. In unweighted networks, the ties are binary: $a_{ij}=1$ if a connection exists and $0$ otherwise, in which case $C_D$ reduces to a simple count of neighbors. Furthermore, in unweighted networks, it is often useful to normalize degree centrality to allow comparisons across nodes and networks of different sizes:
\begin{equation*}
    C_{D}^{'}(n_i) = \frac{1}{N - 1} \sum_{j=1}^{N-1} a_{ij}.
\end{equation*}
Since the maximum number of neighbors is $N-1$, the normalized index ranges between 0 and 1, with 1 indicating a node connected to all others.

The \textit{Eigenvector Centrality} ($C_E$) measure, introduced by \cite{bonacich1987power}, is a measure of structural importance that extends the logic of $C_D$. While degree centrality evaluates the importance of a node based only on the number of its direct ties, eigenvector centrality also considers the quality of those ties; see \cite{bonacich2007some} and \cite{el2021network}. The central idea is that a node is considered more central if it is connected to other nodes that are themselves central. In other words, the centrality of a node depends not only on how many connections it has but also on how well-connected its neighbors are. In practice, $C_E$ assigns to each node $n_i$ a score that is proportional to the sum of the scores of its neighbors, with the weights specified by the adjacency matrix $A$ of the network. The adjacency matrix $A$ is of dimension $N \times N$ in which each entry $a_{ij}$ represents the relationship between node $i$ and node $j$. As in the degree centrality framework, in an unweighted network $a_{ij}=1$ if there is a tie between $i$ and $j$ (and $0$ otherwise), while in a weighted network $a_{ij}$ captures the intensity of the tie, such as the percentage of ownership or the volume of trade. The recursive definition of eigenvector centrality is as follows:
\begin{equation}
    C_E(n_i) = \frac{1}{\lambda} \sum_{j=1}^{N-1} a_{ij} C_E(n_j),
    \label{eq:ec_scalar}
\end{equation}
where $C_E(n_i)$ is the centrality score of node $i$, $\lambda$ is the largest eigenvalue (in absolute value) of $A$, and $a_{ij}$ is the entry of the adjacency matrix $(i,j)$. The key idea is that the importance of a node depends on the importance of its neighbors, whose scores, in turn, depend on their own neighbors. Due to this interdependence, all values of $C_E$ must be solved simultaneously. This becomes possible by rewriting \eqref{eq:ec_scalar} using matrix notation, that is:
\begin{equation*}
    A \mathbf{C_E} = \lambda \mathbf{C_E},
    \label{eq:eig}
\end{equation*}
where $\mathbf{C_E}$ is the eigenvector associated with the largest eigenvalue $\lambda$ of $A$. Each entry in $\mathbf{C_E}$ corresponds to the centrality degree of a particular node. Hence, solving this system yields a consistent set of scores for the entire network, capturing the idea that a node is central if it is connected to other central nodes. These features have made $C_E$ especially useful in contexts where indirect influence matters more than direct connections, such as corporate ownership, international politics, or the diffusion of institutional norms.

\subsection{Brokerage Centrality Measures}
One of the first formulations of brokerage-based centrality is the \textit{Betweenness Centrality} ($C_B$) proposed by \cite{freeman1978centrality}. Intuitively, a node attains high betweenness when it frequently lies on the shortest paths (geodesics) connecting other pairs of nodes. In such a position, the node can facilitate, filter, or control the flow of information, resources, or influence across the network.

Although widely used, for example by \cite{hafner2010centrality} and \cite{goddard2009brokering}, $C_B$ has a well-known structural limitation: it considers only the shortest paths (\cite{newman2005measure}). In many real networks, however, flows do not necessarily follow geodesics but may instead diffuse along multiple, possibly overlapping, alternative routes. To address this limitation, \textit{Flow Betweenness Centrality} ($C_F$) was introduced by \cite{freeman1991centrality} and further developed in subsequent research, including \cite{borgatti2005centrality}, who framed it within a general typology of network flow processes, and \cite{grassi2010vertex}, who applied it to the study of information diffusion in corporate board networks. Unlike $C_B$, $C_F$ evaluates the role of a node in sustaining the maximum flow of resources between pairs of actors. This makes it particularly suited to contexts where flows do not necessarily follow the shortest paths but may instead spread along longer or alternative routes. Conceptually, $C_F$ is based on three principles. First, the capacity of a path depends on its weakest tie: the overall flow is limited by the smallest weight along the path. Second, flows are not restricted to geodesics; actors may rely on longer routes if the shortest ones are blocked or exploitative. Third, a node that channels a significant share of the maximum flow between others has a brokerage advantage, since other actors depend on it to mediate exchanges. Let $m_{jk}$ denote the maximum flow between nodes $j$ and $k$, and $m_{jk}(n_i)$ the portion of that flow that passes through node $i$. The flow betweenness of node $i$ is then defined as:  
\begin{equation*}
    C_F(n_i) = \sum_{j<k, \, i \ne j \ne k} m_{jk}(n_i),
    \label{eq:flow_betweenness}
\end{equation*}  
where the summation covers all distinct pairs of nodes $(j,k)$ that are not equal to $i$. To allow comparability between networks, this measure can be normalized by dividing by the total maximum flow between all pairs where $i$ is not itself an endpoint. By incorporating all possible paths rather than only the shortest ones, $C_F$ provides a more comprehensive picture of the brokerage. It captures not only whether a node is strategically placed on geodesics, but also whether it is indispensable to sustain flows through alternative routes. This broader view makes flow betweenness particularly useful for studying structural hierarchies and coercive power, for example, in the analysis of political alliances or international relations. 

Similarly, the \textit{Walk Betweenness Centrality} ($C_W$), developed by \cite{newman2005measure}, extends the betweenness framework by generalizing the way structural control is measured. Unlike standard betweenness measures, $C_W$ models diffusion as a random-walking process in which information or influence can travel through any possible route connecting two nodes. This shift from a deterministic to a probabilistic representation of flow dynamics offers a more realistic view of how actors can exercise structural control within networks, even through indirect or non-optimal connections. The definition of $C_W$ for a node $n_i$ is:
\begin{equation*}
    C_W(n_i) = \sum_{j \ne i, k \ne i} I_{jk}(n_i),
\end{equation*}
where $I_{jk}(n_i)$ denotes the contribution of node $n_i$ in mediating the connection between $n_j$ and $n_k$. The calculation of $I_{jk}(n_i)$ is based on three matrices: the adjacency matrix $A$, the degree matrix $D$, and the Laplacian $T = D - A$. The degree matrix $D$ is diagonal, with each entry $D_{ii}$ equal to the degree of the node $i$, that is, the number of links in unweighted networks or the sum of weights of the links in the weighted ones. The Laplacian matrix $T$ represents the diffusion structure of the network, while its inverse $T^{-1}$ captures the cumulative influence of all possible paths connecting any two nodes, rather than only the shortest ones. By incorporating $T^{-1}$ into the computation, $C_W$ quantifies how frequently a node mediates flows throughout the network, providing a more general and probabilistic account of brokerage than standard betweenness measures. Empirical evaluations, for example, by \cite{grando2016analysis}, show that $C_W$ offers a high level of detail in distinguishing the structural importance of nodes within a network. It often emerges as one of the most informative indicators of brokerage. However, the main limitation of $C_W$ is its computational complexity: because it requires inverting the Laplacian, the runtime grows rapidly with network size, which makes its application challenging for large networks.

\subsection{Efficiency Centrality Measures}
The \textit{Closeness Centrality} ($C_C$) introduced by \cite{freeman1978centrality} is a global measure of centrality that captures the efficiency and independence of a node within a network. The idea is that a node is more central if, on average, it is closer to all other nodes. This position confers structural advantages in terms of speed, efficiency, and autonomy. Subsequent studies have further examined its conceptual relevance and computational challenges, particularly in large-scale social networks; see, for example, \cite{okamoto2008ranking} and \cite{zhang2017degree}. In particular, these works highlight both the interpretive value of $C_C$ in assessing accessibility within complex structures and the need for efficient algorithms to handle its calculation in very large graphs. 

Nodes with high $C_C$ exhibit three main structural advantages. First, they are more efficient as they can acquire and disseminate resources through fewer intermediaries. Second, they are more independent, since they rely less on others to reach the rest of the network. Third, they achieve faster diffusion, being able to spread information or resources quickly across the system. In contrast, low $C_C$ nodes are structurally distant, slower in accessing or transmitting resources, and more dependent on intermediaries. Formally, the $C_C$ of a node $n_i$ is defined as:
\begin{equation*}
    C_C(n_i) = \left[ \sum_{j \ne i} d(n_i, n_j) \right]^{-1},
    \label{eq:closeness}
\end{equation*}
where $d(n_i, n_j)$ is the geodesic distance between nodes $i$ and $j$. The summation term represents the total distance from node $i$ to all other nodes in the network. A larger sum means that, on average, node $i$ is farther away from the others and therefore less central. In contrast, a smaller sum indicates that node $i$ is closer to all others. To reflect this inverse relationship between distance and centrality, the measure takes the reciprocal of the total distance, so that nodes with shorter average paths obtain higher values of $C_C$.
To allow comparability between networks of different sizes, this measure is usually normalized by multiplying by the maximum possible number of neighbors $(N-1)$:
\begin{equation*}
    C^{'} _C(n_i)= (N - 1) \left[ \sum_{j \ne i} d(n_i, n_j) \right]^{-1}.
\end{equation*}
In this form, the index ranges from 0 to 1, with 1 indicating that the node can directly reach all the others. Despite its intuitive appeal, $C_C$ has lower granularity than other centrality measures, such as $C_E$, meaning that it is less sensitive in distinguishing nodes by their structural importance; for example, \cite{grando2016analysis}). Nevertheless, it remains a valuable indicator for identifying actors that are structurally efficient and independent in accessing or disseminating resources.

The \textit{Information Centrality} ($C_I$) measure introduced by \cite{stephenson1989rethinking} and later applied in works such as \cite{hafner2010centrality} and \cite{grando2016analysis} is a refinement of $C_C$ that offers a more comprehensive measure of structural efficiency. Unlike $C_C$, which relies only on geodesic distance, $C_I$ takes into account all possible paths between nodes, weighting them by their strength and length. In this way, $C_I$ evaluates the efficiency with which information can flow from and to a node throughout the network. Nodes with high $C_I$ are efficient and independent: they can reach others through multiple reliable routes, not just the shortest, making them less dependent on intermediaries. This property makes $C_I$ particularly suitable for contexts where robustness is crucial, for example, when resources, information, or norms diffuse through multiple overlapping routes rather than along a single shortest path. Unlike geodesic-based measures, Information Centrality takes into account the existence of alternative paths connecting the same pair of nodes, weighting their combined contribution to overall connectivity. Starting from the adjacency matrix $A = [a_{ij}]$, a new matrix $X = [x_{ij}]$ of order $n \times n$ is defined as:
\begin{equation*}
    x_{ii} = 1 + \sum_{j \ne i} x_{ij}, \quad x_{ij} = 1 - a_{ij} \quad (i \ne j).
\end{equation*} 
The inverse of this matrix, $B = X^{-1}$, represents the effective distances among the nodes, taking into account both direct and indirect connections. Using this matrix, the Information Centrality of node $n_i$ is computed as:
\begin{equation*}
    C_I(n_i) = \frac{1}{N} \sum_{j=1}^N \frac{1}{b_{ii} + b_{jj} - 2 b_{ij}},
\end{equation*}
where $b_{ij}$ is the element $(i,j)$ of $B$ and $N$ is the number of nodes. This formulation measures how efficiently information can travel between nodes $i$ and $j$ through the entire network. The denominator $(b_{ii} + b_{jj} - 2 b_{ij})$ expresses the effective resistance between two nodes: lower values indicate stronger connectivity. Even in this case, to allow for comparability between nodes, $C_I$ can be normalized:
\begin{equation*}
    C^{\prime}_I(n_i) = \frac{C_I(n_i)}{\sum_i C_I(n_i)}.
\end{equation*}
Its ability to consider multiple paths rather than only geodesics makes it particularly useful for studying diffusion processes, such as information in trade agreements, financial networks, or political alliances (\cite{hafner2010centrality}).

Other measures also fall within these categories but are not discussed here, either because they provide conceptually interesting but, in our opinion, less informative perspectives or because empirical studies have shown that they perform worse than alternative measures in distinguishing the structural importance of nodes. For example, \textit{Eccentricity Centrality} ($C_X$) belongs to the efficiency family and defines the centrality of a node based on the length of the longest geodesic path connecting it to any other node in the network, reflecting the worst-case distance rather than the average. Although conceptually appealing, $C_X$ has been shown to perform worse in empirical applications than $C_C$ and $C_I$, providing lower granularity in distinguishing the structural importance of nodes (\cite{grando2016analysis}).

Centrality measures provide a powerful lens for analyzing network structure; nevertheless, their purely topological nature overlooks the strategic dynamics and decision thresholds that govern corporate control. To address this limitation, power should be understood not only as a matter of structural position but also as the capacity to form winning coalitions, a perspective developed within game-theoretic approaches.

\section{Game-theoretic Measures}
\label{sec:GameTheory}
While centrality measures focus on static topology to identify positional importance, game-theoretic models interpret control as a weighted voting game. In this framework, the influence of an actor does not depend solely on the resources or votes that they hold, but on their ability to form a winning coalition. These models therefore analyze how collective outcomes depend on the relative power of different coalitions. Power is measured as the probability of being pivotal, that is, the likelihood that the participation of an actor is essential to transform a losing coalition into a winning one.

Classical game-theoretic indices, such as the Shapley–Shubik, Banzhaf, and \textit{Johnston index} ($J$-power), assign power scores that capture decisiveness rather than formal allocations of resources or rights. However, these measures were originally designed for simple voting games, and thus struggle to represent the structural complexity of real systems. To overcome these limitations, several extensions have been developed: the Karos–Peters $\Phi$ index generalizes the Shapley–Shubik reasoning to entire networks, while the Mercik–Lobos $\pi$ and $\pi'$ indices refine pivotality to account for cycles and overlapping relationships. These models are particularly useful for analyzing coalition dynamics and control thresholds in complex systems, where indirect alliances often determine who ultimately governs.

However, they face two main challenges. First, they are less intuitive than centrality-based measures, making interpretation and communication more difficult. Second, different indices can assign divergent scores to the same actor, raising questions of comparability and consistency. In this regard, Section~\ref{subsec:ClasIndex} presents the classical indices introduced in cooperative game theory, while Section~\ref{subsec:NetExt} adapts these concepts to the analysis of complex ownership structures.

\subsection{Classical Indexes}
\label{subsec:ClasIndex}
The \textit{Shapley–Shubik index} ($SS$), introduced by \cite{shapley1954method}, is the canonical cooperative game-theoretic measure of voting power and has become a key tool for analyzing corporate control within Complex Network Analysis. Unlike centrality measures, which evaluate structural position, the $SS$ index quantifies the decisiveness of an actor in a weighted voting game: power depends not on the amount of resources held, but on the probability of being pivotal, namely the player whose participation transforms a losing coalition into a winning one. Formally, consider a simple weighted voting game $(N,v)$, where $N={1,\dots,n}$ is the set of players and $v(S)=1$ if a coalition $S \subseteq N$ meets the decision threshold and $v(S)=0$ otherwise. The function $v(\cdot)$ thus defines which coalitions are winning or losing. The index $SS$ of the player $i$ is given by the proportion of all possible voting orders (permutations of $N$) in which $i$ is pivotal:
\begin{equation}
SS(i) ;=; \sum_{S \subseteq N \setminus {i}}
\frac{|S|!,(n-|S|-1)!}{n!},
\bigl[v(S \cup {i}) - v(S)\bigr].
\label{eq:ss}
\end{equation}
\noindent Equation~\eqref{eq:ss} can be understood as the average contribution of the player $i$ to all possible ways in which the players can form a coalition.
For each subset $S \subseteq N \setminus {i}$, the term
$\frac{|S|!,(n-|S|-1)!}{n!}$ gives the probability that the members of $S$ join the coalition before the player $i$ in random order of all the players. The difference $v(S \cup {i}) - v(S)$ checks whether the player $i$ changes the outcome in that situation: it takes the value $1$ when the coalition $S$ is losing but becomes winning once $i$ joins and $0$ otherwise. By adding all possible subsets $S$, the Shapley–Shubik index measures the frequency with which the player $i$ is decisive on average, that is, the frequency with which their participation turns a losing coalition into a winning one.

The \textit{Banzhaf index} ($\beta$), introduced by \cite{banzhaf1964weighted}, is a cooperative game-theoretic measure of voting power that, like the $SS$ index, evaluates how often a player is decisive in a weighted voting game. However, the two measures differ in how they define decisiveness. The index $SS$ considers the order in which players join a coalition and measures the probability of being pivotal in that sequence. In contrast, the $\beta$ index ignores order and focuses only on coalition composition. A player is said to create a swing when their participation turns a coalition from losing to winning, that is, when the coalition including the player is winning ($v(S)=1$), but the same coalition without them would lose ($v(S \setminus {i})=0$). The index $\beta$ therefore counts how many coalitions each player is critical in, assigning equal weight to all possible coalition configurations. Let $\eta_i$ denote the total number of coalitions in which player $i$ is critical, that is, when their participation changes the coalition outcome from losing to winning:
\begin{equation*}
\eta_i = \sum_{S \subseteq N \setminus {i}} [v(S \cup {i}) - v(S)],
\end{equation*}
where the term $[v(S \cup {i}) - v(S)]$ equals $1$ if the addition of player $i$ transforms a losing coalition $S$ into a winning one and $0$ otherwise.
Each coalition in which this condition holds contributes one unit to $\eta_i$, which therefore counts all possible \textit{swings} of the player $i$ throughout the game. The Banzhaf index is then defined as follows:
\begin{equation}
\beta_i(v) = \frac{\eta_i}{2^{n-1}},
\label{eq:banzhaf_abs}
\end{equation}
where the denominator $2^{n-1}$ represents the total number of coalitions that do not include the player $i$, providing the reference set to evaluate all potential swings. The normalized (or relative) Banzhaf index is obtained by dividing the raw value of each player by the sum of all values of all the players, so that the indices collectively sum to one:
\begin{equation}
\beta'_i(v) = \frac{\eta_i}{\sum{j \in N} \eta_j}.
\label{eq:banzhaf_rel}
\end{equation}
\noindent Equation~\eqref{eq:banzhaf_abs} thus expresses how many times, out of all possible coalitions, the inclusion of player $i$ changes the collective outcome from losing to winning. The normalized version in Equation~\eqref{eq:banzhaf_rel} rescales these counts into relative shares, generating a probability distribution of power between players. Although both the $\beta$ and $SS$ indices measure voting power, they differ in how they treat coalition formation: the $\beta$ index gives equal weight to all coalitions, whereas the $SS$ index weights them according to the order in which players join.

The \textit{Johnston index} ($J$-power) introduced by \cite{johnston1978measurement} is a way to measure the influence each player has on a voting system where coalitions can win or lose. The idea is based on two concepts. First, a player is called critical if, without them, a winning coalition would no longer have enough votes to win. Second, a coalition is called vulnerable if it includes at least one such critical player. The index $J$-power assigns one 'unit' of power to every vulnerable coalition and then splits it equally among its critical members. In this way, the index measures how often a player is essential to the success of a coalition, while avoiding giving too much weight to coalitions where several players are decisive at the same time. Let $VC$ denote the set of vulnerable coalitions, and $VC_i$ the subset of those coalitions that include player $i$. The $J$-power index of $i$ is defined as follows:
\begin{equation*}
    J_i(v) = \frac{\sum_{S \in VC_i} r(S)}{\sum_{S \in VC} r(S)} ,
    \label{eq:johnston}
\end{equation*}
where $r(S)=1/k$ if $k$ players are critical in coalition $S$. In this way, $J$-power measures the proportion of total fractional swings that can be attributed to each player. Hence, unlike the $\beta$ index, which simply counts how many times a player is critical across all possible coalitions, the $J$-power index adjusts for situations in which several players are simultaneously critical within the same coalition. Instead of assigning full weight to each of them, as does the index $\beta$, it divides the weight of that coalition equally among all its critical members. In this way, the $J$-power index provides a fairer assessment of individual influence, preventing the total power in a coalition from being overcounted.

It is important to note that these indices share important limitations when applied to corporate control networks. They struggle to capture structural complexities such as cross-shareholdings and external threats because they are designed for single, isolated voting games rather than interconnected or dynamic systems. This limitation motivates the next section (\ref{subsec:NetExt}), which introduces methodologies that extend these indices to complex networks.

\subsection{Extension of Classical indices}
\label{subsec:NetExt}
The \textit{Karos--Peters index} ($\Phi$), introduced by \cite{karos2015indirect}, extends the $SS$ logic of pivotality to networked structures with cross-shareholdings. It is designed to capture indirect forms of control and to assess the overall contribution of each actor to decision-making within an interconnected system. The approach begins with the definition of a mutual control invariant structure $C$, which formalizes the control relationships among actors in the network.
Consider a set of actors $N$ and any subset $S \subseteq N$ representing a coalition of controlling entities.
The function $C(S)$ denotes the set of actors that are effectively controlled by $S$, either directly through ownership links or indirectly through other entities already under the control of $S$. 

The $\Phi$ index applies the Shapley--Shubik index to each of these control games to evaluate the contribution of each firm to the control of all others in the network. In particular, for each $i \in N$, the index is defined as:
\begin{equation}
\Phi_i(C) = \sum_{k \in N} \left[SS_i\bigl(v_k^C\bigr)- v_i^C(N)\right],
\label{eq:phi}
\end{equation}
where $SS_i(v_k^C)$ denotes the Shapley–Shubik value of actor $i$ in the control game that determines whether a coalition can control actor $k$; and $v_i^C(N)=0$ for any player $i$ who is never required by any coalition to exert control and who is not controlled by any coalition. The sum of all $k \in N$ provides the general frequency with which the actor $i$ is pivotal in the various control games that make up the network. The term $v_i^C(N)$ is subtracted to remove the trivial case of self-control, since an actor is automatically controlled when the coalition includes all players. Hence, with \eqref{eq:phi} it is possible to evaluate both the ultimate and intermediate controllers within the same analytical framework. However, a key limitation of $\Phi$ is that it often reproduces the initial distribution of resources rather than captures how these resources translate into effective decision-making influence; see \cite{mizuno2023flow}.

The \textit{Mercik--Lobos} ($\pi$) index, also known as the implicit power index, was introduced by \cite{mercik2016index} to measure direct and indirect control in networks with reciprocal relations and feedback loops. Unlike $\Phi$, which is based on the $SS$ index, $\pi$ is derived from the $J$-power. As argued by \cite{mercik2016index}, $J$-power is better suited to highly interdependent systems, where actors are tightly connected and the network behaves as a unified structure. In such contexts, even small changes in one connection can cascade through the entire network. Building on this idea, the $\pi$ index applies the logic of $J$-power to network settings by treating the control of each entity as a simple game. Its computation involves three steps: (i) calculating absolute Johnston values for all participants and groups, (ii) redistributing the power of collective actors equally among their members, and (iii) summing and normalizing these contributions across the system. The resulting score $\pi_i$ represents the implicit power of the actor $i$ within the network. Originally conceived to address complex topologies such as feedback cycles and two-link mutual control structures, the $\pi$ index nevertheless faces two main criticisms. First, it often reflects how much an actor is controlled by others rather than how much control it actually exercises, making it poorly suited to capture the role of intermediaries, since final actors may obtain higher scores than strategically placed middle actors. Second, the original $\pi$ index does not satisfy the property of a null player, meaning that even actors who are never critical in any coalition may still receive positive power.

To address this limitation, \cite{mercik2018measurement} proposed a modified version, denoted $\pi'$. The crucial change is in the redistribution step: while $\pi$ divides the power of collective actors equally among their members, $\pi'$ allocates it proportionally to their absolute Johnston values. This correction ensures that null players receive zero power, restoring a desirable property that the original $\pi$ failed to achieve.

\section{Concentration Measures}
\label{sec:conc}
Concentration measures provide a way to assess the uneven influence or distribution of resources within a system. Rather than examining the role of individual actors, they capture how much outcomes are dominated by a small subset of participants. In this framework, concentration is evaluated by comparing the overall distribution in the system with the portion accumulated by the largest players. A system is highly concentrated when a very small group holds a disproportionately large portion of the total. This approach considers not only direct endowments but also how influence can accumulate indirectly through interconnected relationships. Compared to the approaches previously shown, concentration measures emphasize accumulation: how much influence or resources end up clustered in the hands of a few. In this sense, they complement centrality and game-theoretic approaches, providing an aggregate view of inequality in distribution that cannot be captured by positional or coalition-based logics alone. This section defines the most relevant indices used in this framework.

The \textit{Herfindahl--Hirschman index} ($HHI$) is a classical measure of concentration that captures how unevenly a quantity is distributed across the elements of a system (\cite{hirschman1980national}). It is computed by squaring the proportion assigned to each actor and summing across all actors, so that larger contributions weigh disproportionately more than smaller ones. It is defined as
\begin{equation}
    HHI = \sum_{i=1}^n p_i^2 ,
    \label{eq:hhi}
\end{equation}
where $p_i \in [0,1]$ is the share of the total held by the actor $i$, with $\sum_{i=1}^n p_i = 1$. The index ranges from $HHI=1$ when a single actor holds everything to values approaching 0 when the total is evenly distributed among many actors. The strength of $HHI$ lies in its simplicity, providing a simple aggregate view of concentration, which is useful to identify when a small set of actors dominates a system. However, as an aggregate index, it does not capture the structural positions or coalition dynamics that centrality and game-theoretic measures are designed to reveal.

The \textit{Top-k ratio} index is a straightforward measure of concentration that captures how much of the total is held collectively by the $K$ largest actors; see \cite{hirschman1980national}. In particular, it is defined as:
\begin{equation*}
    Top_k = \sum_{i=1}^K p_i,
\end{equation*}
where $p_i$ are defined as in \eqref{eq:hhi}, but are now ordered such that $p_1 \geq p_2 \geq \dots \geq p_n$. The measure simply adds up the proportions of the top $k$ actors, showing how dominant this elite group is relative to the whole. Compared to $HHI$, which squares each share and is therefore sensitive to the full distribution (giving more weight to large actors), the Top-k ratio focuses only on the cumulative dominance of a few at the top. Hence, it highlights elite concentration, but says little about how power is distributed among the rest of the actors. Its appeal lies in its immediacy: the ratio $Top-k$ is easy to interpret and widely used in policy and antitrust contexts to assess whether a small set of actors exercises disproportionate influence. However, it has two main drawbacks. First, the choice of $k$ is arbitrary and may change the interpretation of the results. Second, like the classical $HHI$, it captures only direct shares and ignores the deeper structure of the network, such as indirect ties or multilevel relationships.

The \textit{Net Control index} (NCI), introduced by \cite{brancaccio2018centralization}, aims to quantify capital centralization through a methodology explicitly grounded in network analysis. Unlike classical concentration measures, it captures not only direct ownership but also the indirect links that propagate control across chains of corporate relations. 
Specifically, NCI was defined as the proportion of shareholders that together represent at least a fixed share $H$ of total control. More precisely, let $N$ be the total number of shareholders and let $C_i$ denote the control share of shareholder $i$, with all $C_i$ ranked in descending order. Then, it is possible to identify the smallest group of actors $G_{\gamma}$ such that the sum of their control shares reaches or exceeds the chosen threshold $H$ (e.g. $80\%$ of the market). The index is given by:
\begin{equation*}
g_{\gamma} = \frac{|G_{\gamma}|}{N} \times 100,
\end{equation*}
where $|G_{\gamma}|$ is the number of shareholders in this minimal elite. In other words, $g_{\gamma}$ expresses what fraction of all shareholders is sufficient to command a dominant cumulative share of control. A lower value of $g_{\gamma}$ indicates a higher concentration, since it implies that only a very limited group of actors can reach the threshold $H$. Using this method, \cite{brancaccio2018centralization} empirically shows that, at the global level, less than $2\%$ of the shareholders are sufficient to control more than $80\%$ of the listed capital, highlighting the extreme centralization of ownership structures. However, like the other concentration indices discussed in this section, it does not provide information about how control is distributed within that group, nor about the role of intermediate shareholders that transmit control without necessarily belonging to the final elite. As a result, it is best interpreted as a measure of overall centralization, rather than as a full description of networked power relations.

Although the \textit{Ultimate Control} (UC) method, introduced by \cite{la1999corporate}, is not a concentration measure like $HHI$ or NCI, it provides a core metric for identifying how power concentrates in the hands of ultimate owners. For this reason, it is classified here among concentration-based methodologies, as it supports the structural analysis of how control is consolidated within corporate and institutional networks. The method identifies the ultimate controller by tracing ownership chains through direct and indirect participations, applying a deterministic principle of transitivity. An entity is deemed ultimately controlled when a shareholder (whether an individual, family, corporation, or public authority) directly or indirectly exceeds a fixed voting percentage threshold, for example, $20\%$. The analysis proceeds recursively until reaching an actor not controlled by any other, thus defining the ultimate owner. This logic captures both direct and indirect control and shows how pyramidal structures amplify influence while requiring limited capital. Whereas centrality measures capture structural position and game-theoretic models view power as pivotality within coalitions, the UC method defines control in legal and institutional terms, identifying formal authority rather than functional influence. Despite its clarity and wide use (see, for example, \cite{lin2021varieties} and \cite{dong2022does}), its reliance on fixed thresholds produces a static and hierarchical view of power, overlooking the role of minority shareholders and intermediaries that may exercise influence in fragmented or multi-layered systems.

\section{Flow-Based Measures}
\label{sec:Flow}
Flow-based measures evaluate how resources, influence, or control propagate through the links of a network. Rather than focusing only on the static position of actors, they conceptualize influence as a dynamic flow that spreads along relational chains and typically attenuates as the paths become longer. The key idea is that the importance of a node depends not only on its direct links, but also on the cumulative contribution of multiple indirect pathways. Hence, flow-based measures provide a dynamic perspective on power: they capture how influence diffuses and accumulates across direct and indirect connections, while accounting for attenuation along long chains. Here, we focus on the most relevant approaches within this framework.

The Network Control Value ($NCV$) was introduced by \cite{vitali2011network} to quantify the amount of control transnational corporations exert both directly and indirectly; see also \cite{rungi2017global}. Their analysis revealed a bow-tie structure in the global corporate ownership network, where a small and densely interconnected core of transnational corporations holds a disproportionate share of global control. In particular, it was identified about 737 top holders, roughly 0.6\% of all shareholders, accumulating nearly 80\% of the total operating revenue of transnational corporations, most of them being large financial intermediaries located in the core of the network. This finding is consistent with the evidence reported by \cite{brancaccio2018centralization}, who, using the NCI framework discussed above, show that less than 2\% of global shareholders control around 80\% of the listed capital of the world. Although NCV was originally developed for corporate ownership networks, its methodology has a much broader scope. Because it measures how a scalar quantity propagates through a weighted and directed network, it can also be applied to the study of political and social influence (diffusion of norms and standards), systemic risk in financial networks, and global supply chains (tracing how shocks cascade across production stages). Formally, NCV quantifies the total value influenced by an actor, accounting for both direct and indirect ties, and treating the influence as a flow that propagates through the network along weighted and directed connections. For example, if firm $A$ controls $B$, and $B$ controls $C$, then part of the influence of $A$ extends to $C$. This recursive logic is captured in the definition of the NCV of the actor $i$:
\begin{equation}
\label{eq:NCV}
NCV_i = \sum_j C_{ij} v_j + \sum_j C_{ij} NCV_j,
\end{equation}
The first term, $\sum_j C_{ij} v_j$, captures direct control: it measures the portion of node $j$’s value or capacity—such as resources, output, or influence—that is directly affected by actor $i$, weighted by the control share $C_{ij}$ linking $i$ to $j$.
where the first term, $\sum_j C_{ij} v_j$, captures direct control: it measures the portion of node $j$’s value or capacity (such as resources, output, or influence) that is directly affected by actor $i$, weighted by the control share $C_{ij}$ linking $i$ to $j$. The second term, $\sum_j C_{ij} NCV_j$, captures indirect control: since $NCV_j$ already includes the direct and indirect influence of $j$, controlling $j$ means that $i$ inherits a share of everything $j$ controls. 
However, in spite of its importance to capture the total influence an actor exerts (including indirect ties), the NCV has inherent weaknesses. In networks with cycles or cross-shareholdings, the recursive propagation of control can loop indefinitely, leading to double-counting and an overestimation of the influence of certain actors. This issue was one of the main reasons that led to the introduction of the net NCV.

The \textit{net Network Control Value} ($nNCV$), introduced by \cite{vitali2011network}, is a refinement of the original $NCV$. As discussed previously, the $NCV$ framework tends to overestimate influence in the presence of ownership and cross-shareholding cycles, where control can circulate and be counted multiple times. The $nNCV$ corrects this bias by distinguishing the portion of control that an actor genuinely generates from that which is merely recycled through such loops. It is defined as:
\begin{equation*}
    nNCV_i = NCV^{\mathrm{P}}_i - NCV^{\mathrm{R}}_i,
\end{equation*}
where $NCV^{\mathrm{P}}_i$ is as defined in \eqref{eq:NCV}, while $NCV^{\mathrm{R}}_i$ captures the share of control that $i$ does not generate autonomously but only inherits passively from others through circular ownership. Subtracting $NCV^{\mathrm{R}}_i$ eliminates artificial inflation caused by ownership cycles while preserving direct and legitimate indirect effects. Although $nNCV$ corrects the main weakness of the original $NCV$, it still exhibits important limitations. Its reliance on fixed control thresholds implies that actors holding less than a majority stake are treated as having no influence, thereby overlooking the potential role of minority shareholders and intermediaries that may be decisive in channeling or brokering control.

Flow-based centrality measures, such as \textit{PageRank} (PR) and \textit{Katz Centrality}, can complement nNCV by capturing how influence propagates along both direct and indirect paths within a network, offering a more nuanced view of structural importance. They were originally introduced as centrality measures in graph theory: $PR$ as an algorithm to rank web pages and Katz Centrality as a classical centrality index. However, in the study of complex networks, both have been successfully reinterpreted as flow-based measures. They capture how influence is redistributed along direct and indirect paths, offering a structural perspective on the way dispersed resources, information, or control propagate through a network.

$PR$ defines recursively the score of node $v_i$ as:
\begin{equation*}
    PR(v_i) = (1 - \alpha) + \alpha \sum_{v_j \in In(v_i)} \frac{PR(v_j)}{Out(v_j)},
\end{equation*}
where $\alpha \in (0,1)$ is a damping factor (commonly $\alpha = 0.85$) that reduces the weight of long propagation paths and guarantees convergence to a unique equilibrium. In this formulation, $In(v_i)$ denotes the set of nodes that link to $v_i$, while $Out(v_j)$ represents the number of outgoing connections from node $v_j$. Each node distributes its PageRank value evenly across all the nodes it points to, so that every outgoing edge transmits only a fraction $\tfrac{PR(v_j)}{Out(v_j)}$ of its total influence. Intuitively, a node is important if it is linked to by other important nodes, but its received influence is diluted when those nodes spread their connections across many others. The strength of the measure lies in its simplicity and its ability to capture both direct and indirect connectivity in a unified recursive framework. The main limitation of PageRank is that it assumes that a node is important just because many others point to it. In reality, having many links does not always mean having real power: to influence decisions, what matters is whether those links give enough votes to pass a majority threshold or allow the node to play a decisive role in coalitions. PageRank does not capture this, so a node may look very important in the ranking even if it cannot actually control outcomes.

Katz Centrality follows a similar recursive logic to PageRank, but the two measures differ in both their mathematical formulation and their conceptual goal. PageRank distributes the importance of a node $v_j$ equally among all its outgoing links, so that the score of a node $v_i$ depends on the normalized flow it receives from its direct neighbors. In contrast, Katz Centrality explicitly incorporates the contribution of all possible paths in the network, not just the immediate ones. The cumulative influence matrix is defined as:
\begin{equation*}
    T = \sum_{l=1}^{\infty} \alpha^{\,l-1} A^l \;=\; A (I - \alpha A)^{-1},
\end{equation*}
where $A$ is the adjacency matrix (or direct control) and $A^l$ captures all walks of length $l$. The attenuation factor $\alpha$ reduces the weight of long paths, so that the measure balances direct control with indirect path-dependent influence. If $\alpha = 0$, the index collapses to direct control ($T = A$); as $\alpha \to 1$, the measure approaches purely transitive influence, where paths of all lengths contribute almost equally. This formulation captures the idea that total influence is the sum of all possible ways in which control can propagate through the network. Direct links contribute fully, while indirect links contribute with decreasing weight as the paths become longer. Rewriting the infinite series in closed form, $A (I - \alpha A)^{-1}$, guarantees convergence and makes the computation feasible. 

A more recent development inspired by this formulation is the $\alpha$--ICON algorithm, introduced by \cite{polovnikov2021alpha}. Although it relies on the same mathematical structure as Katz Centrality, its conceptual goal and interpretation are fundamentally different. Katz Centrality uses the attenuation factor $\alpha$ to measure how influence or visibility spreads through a generic network, producing a single score for each node that reflects its overall importance within the system. In contrast, $\alpha$--ICON applies the same recursive logic to corporate ownership data, interpreting the entries in the adjacency matrix $A$ not as abstract connections, but as weighted equity links. Hence, in this case, the resulting matrix $T$ does not measure social prominence, but the cumulative control that a shareholder $i$ exercises (both directly and indirectly) over other firms $j$. By modeling control as a flow that attenuates with distance, $\alpha$--ICON bridges network theory and corporate governance analysis, providing a convergent and interpretable framework for identifying ultimate controlling entities in large global ownership networks.

In general, the main limitation of flow-based measures is that they treat a node as important simply because it receives a large volume of incoming links. However, large inflows of influence do not automatically translate into real power. In many contexts, what ultimately matters is whether these connections enable an actor to cross a critical threshold or to be pivotal in shaping collective outcomes. As a result, a node can appear highly ranked, whether because it channels substantial resources, information, or influence through multiple pathways, but it still lacks the capacity to determine results. To address this limitation, the Network Power Index (NPI) and Network Power Flow (NPF) combine flow-based propagation with game-theoretic logic, embedding the pivotal role of actors within majority thresholds. They thus extend the structural intuition of PageRank- and Katz-like models, but translate influence flows into effective decision-making power. As they merge game-theoretic and flow perspectives, they are considered hybrid approaches and discussed in Section~\ref{sec:Hybrid}.

Finally, although not a flow-based model in the modern probabilistic sense, the approach introduced by \cite{brioschi1989risk} represents an early and rigorous attempt to formalize how interdependent quantities circulate within networked systems. Its use of matrix algebra and recursive calculations of integrated relationships make it a direct precursor of flow-oriented methodologies such as NCV mentioned above. The approach is based on a simple but powerful intuition: the state or influence of each entity depends not only on its own characteristics but also on those of the entities to which it is connected. Each link transmits a fraction of this quantity, whether it is ownership or another form of control, which in turn feeds back through the network as connected entities interact with others. By iterating this process, the model captures the influence that each participant posses, once all direct and indirect relationships are taken into account.

\section{Optimization Measures}
\label{sec:Optim}
Optimization measures study power in networks by determining the smallest group capable of dominating the system and the nodes that are indispensable for maintaining control. In other words, rather than describing connections or tracing flows, they compute the most efficient configuration of actors needed to reach a target. In a corporate ownership network, for example, they can scan all combinations of shareholders and identify the minimal set that jointly controls 80$\%$ of the system, providing a maximum concentration target. The same logic also reveals strategic intermediaries that, despite limited stakes, are essential because influence must pass through them. The strength of this approach lies in its prescriptive nature: optimization models show the minimum structure required to maximize control.

The \textit{Indirect Control} (IC) model, introduced by \cite{martins2017corporates}, frames the problem of gaining power in ownership networks as a mixed-integer optimization task. Its objective is to determine the cheapest way to bring a set of target nodes $T \subseteq N$ under control, combining direct acquisitions with the cascading effects of indirect links in the network $G=(N, A)$. Even in this case, $N$ is the set of all nodes (for example, firms, organizations, or actors), and $A$ is the set of arcs (directed links) between them. Each edge $(i,j) \in A$ has a weight $s_{ij}$, representing the fraction of $j$ controlled by $i$. For every target $j \in T$, with $ T\subseteq N$, a minimum threshold $\alpha_j$ must be met for $j$ to be considered controlled. Direct acquisitions are represented by continuous variables $z_j \ge 0$, while binary variables $x_j \in \{0,1\}$ indicate whether non-target nodes $j \in N \setminus T$ are brought under control. Each direct acquisition has an associated unit cost $p_j$, so the optimization minimizes the total cost:
\begin{equation*}
    \min \sum_{j \in N} p_j z_j.
\end{equation*}
The key to the model lies in the control constraints, which differ for target and non-target nodes. For non-targets $j \in N \setminus T$, the constraint ensures that if $j$ is marked as controlled ($x_j=1$), then the threshold $\alpha_j$ is met by combining direct purchases and indirect contributions from already controlled predecessors:
\begin{equation*}
    \alpha_j x_j \;\le\; z_j + \sum_{i\in\delta^-(j)\cap (N \setminus T)} s_{ij} x_i + \sum_{i\in\delta^-(j)\cap T} s_{ij}, 
    \quad \forall j \in N \setminus T.
\end{equation*}
For targets $j \in T$, the threshold must be satisfied unconditionally:
\begin{equation*}
    \alpha_j \;\le\; z_j + \sum_{i\in\delta^-(j)\cap (N \setminus T)} s_{ij} x_i + \sum_{i\in\delta^-(j)\cap T} s_{ij}, 
    \quad \forall j \in T.
\end{equation*}
These constraints capture how control propagates: it can come from direct acquisitions ($z_j$), but also from the shares inherited through chains of already controlled intermediaries. Solving the model under these conditions identifies both the minimal set of acquisitions needed and the pathways through which control spreads most efficiently. Extensions such as IC2 and IC3 refine the constraints further by excluding feedback from targets or reciprocal cross-shareholdings, thereby avoiding artificial inflation of control. 
 
Despite its usefulness, the IC model has structural weaknesses. The most important is the systematic overestimation of control in the presence of cycles and cross-shareholdings. Because IC allows reciprocal ownership loops, optimization can assign control to an actor even when it depends on circular or fragile structures. This circular dependency inflates the measured influence and can make the solution ambiguous, suggesting that control can be achieved with unrealistically small investments. Moreover, IC allows the use of shares held by target nodes to justify control over other nodes, further exaggerating the effective reach of influence. These limitations motivated the introduction of the Cheapest Control Problem ($CCP$) by \cite{di2019corporate}, a refined formulation that corrects overestimation by preventing circular feedback and preventing the purchase of significant stakes at market price. In doing so, $CCP$ provides a more reliable and economically plausible reference for indirect control.

Although the IC and CCP models provide valuable benchmarks for analyzing indirect influence in ownership networks, both approaches still face important methodological limitations. First, they are static and deterministic: they compute the least-cost path to control based on a single snapshot of ownership and prices, overlooking the dynamic nature of corporate structures, where shares are traded, prices fluctuate, and control strategies evolve. Second, they minimize acquisition cost given the network topology, but disregard legal and institutional constraints, such as takeover thresholds, mandatory public offers, and antitrust rules, as well as strategic reactions from other actors who may resist or counter acquisitions.

\section{Hybrid Measures}
\label{sec:Hybrid}
The approaches analyzed so far usually focus on one family of methods at a time. Each of these lenses provides valuable insights, but they are also partial: structural centrality ignores the dynamics of propagation, flow measures describe diffusion but not strategic costs, and optimization highlights efficiency but not the role of position. Hybrid models combine these perspectives to overcome such limits. By integrating different methodological families, they can reveal information otherwise difficult to capture. Here, we investigate two of the most important hybrid approaches that integrate game theoretic with flow-based measures.

The \textit{Network Power Index} (NPI) combines the game-theoretic logic of the $SS$ index with the flow-based representation of influence networks, modeling how control propagates through chains of direct and indirect connections. In other words, the NPI of a node measures the probability of being pivotal once the transmission and aggregation of influence along relational paths are considered. Because evaluating all possible coalitions across a large and interconnected network would require an exponential number of computations, the exact calculation of the NPI is practically infeasible. Therefore, the index is estimated through an iterative sampling procedure. At each iteration, the nodes are randomly ordered and their associated weights are added sequentially until the cumulative total crosses the decision threshold (for example, 50$\%$). The node that causes the threshold to be reached is identified as the pivot, since without it the coalition would fail, whereas with it the coalition succeeds. Once the pivotal node is determined, its influence is traced upstream and downstream through network connections to reconstruct the control path it activates. In this way, the algorithm captures the entire chain through which decisive power propagates across the system. In a single iteration $t$, after sampling a control structure, the downstream influence vector is computed as
\begin{equation*}
\tilde{y}^{(t)} = [I - d\,Y^{(t)}]^{-1} v,
\label{eq:npi-single}
\end{equation*}
where $Y^{(t)}$ is the control matrix generated in that simulation, whose links are determined by the pivotal paths identified in iteration $t$;  
$d \in (0,1)$ is a damping factor that ensures convergence and discounts long or cyclical propagation paths; and  
$v$ is the vector of initial endowments, representing exogenous resources or rights before propagation. In other words, $Y^{(t)}$ captures the specific configuration of the control links resulting from the random coalition formation process. Multiplying $[I - dY^{(t)}]^{-1}$ by $v$ traces how these initial endowments propagate through the network along direct and indirect control paths in this realization. Repeating this for $T$ simulations yields the estimator obtained by averaging:
\begin{equation*}
\hat{y} \;=\; \frac{1}{T} \sum_{t=1}^{T} \tilde{y}^{(t)} \;=\; \frac{1}{T} \sum_{t=1}^{T} (I - d\,Y^{(t)})^{-1} v,
\label{eq:npi-average}
\end{equation*}
which provides the expected (average) level of power transmitted by each node across randomly realized control structures. The NPI provides a robust measure of decision-making power. However, its scope is limited because it does not capture the role of intermediate actors in control chains. By construction, the NPI focuses on the probability that an ultimate controller becomes pivotal in forming a winning coalition. This design highlights the power of ultimate controllers but overlooks the role of intermediaries that channel and aggregate influence, enabling control to consolidate across the network. The NPI, therefore, fails to capture the contribution of actors such as large asset managers or subsidiaries, which, despite lacking ultimate control, significantly shape how power propagates through interconnected systems. This limitation has motivated the development of the \textit{Network Power Flow} (NPF), introduced by \cite{mizuno2023flow}, which addresses this limitation by explicitly modeling the flow of control along ownership chains. In this framework, intermediaries are explicitly evaluated according to how often they act as critical hubs or pivotal bridges along ownership paths. 

The NPF offers a more comprehensive view than the NPI by capturing both the decisional power of the ultimate controllers and the role of intermediaries in transmitting influence across the network. In each iteration $t$, the control matrix $Y^{(t)}$ is generated by simulating coalition formation and identifying pivotal nodes. Unlike the NPI, which aggregates the results into a control probability for the ultimate controllers, the NPF weights the downstream influence by the endowment of each target and averages the results in simulations:
\begin{equation*}
\hat{p}_{ij}
=
\frac{1}{T}\sum_{t=1}^{T} \tilde{y}^{(t)}_{ij}\, v_j .
\label{eq:npf-avg}
\end{equation*}
Here, $\tilde{y}^{(t)}{ij}$ measures the share of control flowing from actor $i$ to actor $j$ in simulation $t$, while $v_j$ denotes the value associated with $j$ (e.g., its size or voting rights). Multiplying $\tilde{y}^{(t)}{ij}$ by $v_j$ and averaging over all iterations yields the estimator $\hat{p}{ij}$, which represents the expected amount of value transmitted from $i$ to $j$. This formulation makes explicit the contribution of intermediaries: even without ultimate control, their position along control paths determines how much of the’ value of $j$ is mediated through them. Once all $\hat{p}{ij}$ are collected in the matrix $\hat{P} = [\hat{p}_{ij}]$, this matrix serves as the basis for further analysis. Comparing the distribution of $\hat{P}$ with structural indicators, such as $HHI$, allows us to assess whether control is concentrated or dispersed and reveal the functional role of intermediaries in consolidating fragmented resources into effective decision-making power.

\section{Conclusions: Comparative Insights and Research Perspectives}
\label{sec:Conclusions}
Over time, several indices have been developed to quantify power, control, and influence within complex ownership networks. However, these methodologies evolved along different theoretical trajectories, each grounded in a specific interpretation of what “power” represents and how it operates through network structures. As a result, the literature has accumulated valuable tools, but lacks a unified framework that connects their conceptual foundations and clarifies which dimension of control or influence each measure captures. To fill this gap, we propose a taxonomy that classifies the main methodologies according to the conceptual logic through which they define and measure structural power, rather than by their technical features or data requirements. Specifically, we identify six main categories: centrality-based measures conceptualize power as a function of position and connectivity; game-theoretic approaches interpret it as the ability to influence collective decisions; concentration measures define it through the unequal distribution of control; flow-based methodologies describe it as a process that propagates through ownership or relational chains; and optimization models frame it as the outcome of strategic behavior aimed at minimizing control costs. Finally, hybrid approaches combine elements from these perspectives into a single analytical framework.

The qualitative assessment reported in Table \ref{tab:comparison} is based on a comparative evaluation of the ability of each methodology to capture two key analytical dimensions: (i) identification of ultimate controllers, that is, actors who effectively dominate decision-making at the top of ownership or influence chains, and (ii) evaluation of intermediary power, referring to actors that channel, aggregate or broker control between otherwise disconnected parts of the network.
Each method was assigned a qualitative score (Excellent, Good, Fair, or Poor), reflecting its theoretical reach, empirical support in the literature, and its ability to capture the two structural dimensions of power. These evaluations reflect how each methodological family conceptualizes and translates control into measurable network outcomes. For instance, centrality measures such as Degree, Eigenvector, and Closeness describe how visible or well-connected an actor is within the network, but they provide little information about actual control. In contrast, Betweenness-type centrality indices perform better in identifying intermediaries, as they capture brokerage positions that mediate the flow of ownership or influence between otherwise disconnected actors. Flow-based methodologies such as PageRank, Katz Centrality, and $\alpha$–ICON extend centrality measures by modeling how influence and control propagate through ownership chains. These measures are conceptually strong for identifying ultimate controllers, since they capture the transitive accumulation of influence along multiple layers of ownership. However, their ability to detect intermediary power remains limited, as they emphasize end-point accumulation more than the role of nodes that transmit control. Game-theoretic approaches, including the $SS$, Banzhaf and $J-$power indices, view power as decisional influence within collective structures, capturing the probability of being pivotal in coalition formation. Their network-level extensions ($\Phi$, $\pi$, $\pi'$) improve this logic by considering cross-shareholdings and cycles, but they remain less suited to describing intermediation because pivotality is defined within discrete decision games rather than continuous control flows. In this regard, hybrid methodologies combine the decisional logic of game theory with the propagation mechanisms of flow-based models. In this framework, NPI performs well in tracing ultimate controllers, as it quantifies the probability that an owner is decisive across interconnected firms, while the NPF complements it by focusing also on intermediaries, modeling how control originating from ultimate owners diffuses through the network. Concentration-based measures, such as $HHI$ or Top-$k$, remain useful for summarizing how control is distributed at the system level, but provide little information about its transmission or intermediation. Finally, optimization-based models, including IC and CCP, evaluate control as the outcome of strategic decisions that minimize the cost of acquiring influence, capturing both the structural and functional aspects of network ownership with moderate accuracy in both dimensions. Together, the methodologies reviewed here should not be seen as substitutes, but as complementary analytical tools. Each family emphasizes a distinct mechanism (positional, flow-based, decision-based, concentrative, or strategic), and their combined use allows a richer understanding of how control is accumulated, transmitted, and exercised within complex ownership networks.

A promising direction for future research is to refine the game-theoretic foundation of the NPF by replacing the Shapley–Shubik notion of pivotality with the $J$-power measure. As discussed in Section~\ref{sec:GameTheory}, $J$-power offers greater axiomatic consistency and models complex ownership structures with cross-shareholdings and cycles more effectively than the Shapley–Shubik index. Unlike SS, which defines power solely as the probability of being pivotal, $J$-power redistributes influence among all decisive actors within each winning coalition, accounting for joint control situations. Incorporating this framework would make the NPF more robust in densely interconnected networks, where a purely pivotal view may overstate the dominance of ultimate owners and overlook the influence of intermediaries.

\begin{table}[H]
\centering
\resizebox{\textwidth}{!}{%
\begin{tabular}{lccp{9cm}}
\toprule
Method & Ultimate Controllers (UC) & Intermediary Power (IP) & Underlying assumption \\ \\
\midrule
\textbf{Degree, Eigenvector, Closeness} & Poor & Poor & Power as visibility and proximity: influence from direct ties, importance of connected nodes, or minimal distance from others (\textbf{Centrality}). \\ \\
\textbf{Betweenness, Flow-Betweenness} & Fair & Good & Power as brokerage: control over flows by being on shortest paths or maximum-flow routes (\textbf{Centrality}). \\ \\
\textbf{PageRank} & Fair & Good & Flow of importance/capital: influence propagates through the network, discounted over longer paths (\textbf{Flow-based}). \\ \\
\textbf{Katz Centrality, $\alpha$-ICON} & Excellent & Fair & Flow of importance/capital: influence propagates through the network, discounted over longer paths (\textbf{Flow-based}). \\ \\
\textbf{SS, Banzhaf, J} & Fair (single firm) & Poor & Voting power: probability of being pivotal in forming a majority coalition within one firm (\textbf{Game-theoretic}). \\ \\
$\boldsymbol{\Phi, \pi, \pi'}$ & Good & Fair & Network-level decisiveness: game-theoretic power indices extended to networks, handling cycles and cross-holdings (\textbf{Game-theoretic}). \\ \\
\textbf{HHI, Top-$\boldsymbol{k}$} & Good & Poor & Concentration of control: cumulative share of top-$k$ owners or overall inequality (\textbf{Concentration}). \\ \\
\textbf{NCV, nNCV} & Good & Fair & Total control volume: direct and indirect control accumulated through ownership chains (\textbf{Flow-Based}). \\ \\
\textbf{NPI} & Excellent & Fair & Probability that an ultimate controller is pivotal in forming a coalition across the network (\textbf{Hybrid}). \\ \\
\textbf{NPF} & Good & Excellent & Influence on control flow: intermediaries’ impact on the control originating from ultimate controllers (\textbf{Hybrid}). \\ \\
\textbf{IC, CCP} & Good & Good & Cost-minimizing strategies: identify the least costly investment paths to gain control (\textbf{Optimization}). \\ \\
\bottomrule
\end{tabular}}
\caption{\textit{Comparison of key network power measures examined in this study. The evaluation scale is qualitative: Excellent = best suited and widely validated; Good = appropriate but with some limitations; Fair = partially informative, useful only in specific contexts; Poor = not suited for the dimension considered.}}
\label{tab:comparison}
\end{table}

\bibliographystyle{unsrtnat}  
\bibliography{sample}

\end{document}